\documentclass[%
 reprint,
 amsmath,amssymb,
 aps,
]{revtex4-2}

\usepackage{graphicx}
\usepackage{dcolumn}
\usepackage{bm}

\begin{document}

\preprint{APS/123-QED}

\title{Quantum Stoner-Wohlfarth model of two-dimensional single domain magnets}

\author{Essa M. Ibrahim}

\author{Shufeng Zhang}%

\affiliation{%
Department of Physics, University of Arizona, 1117 E 4th Street, Tucson, AZ 85721}%

\date{\today}

\begin{abstract}
The Stoner-Wohlfarth (SW) model is a classical model for magnetic hysteresis of single domain particles. For two dimensional magnets
at finite temperature, the SW model must be extended to include intrinsic strong spin fluctuations. We predict several fundamentally different hysteresis properties between 2D and 3D magnets.
The magnetization switching diagram known as the Asteroid figure in the conventional SW model becomes highly temperature dependent and asymmetric with respect to the transverse and longitudinal magnetic fields. Our results provide new insights for 2D magnetic materials based spintronics applications. 
\end{abstract}

\maketitle

\section{Introduction}
In the last several years, many 2D magnetic materials with novel magnetic and spin transport phenomena have been
discovered \cite{Huang,Gong,Ohara,Klein,Jiang, Lee, Huang2,Gong2, Song,Gupta, MacNeill, Alghamdi,Wang,Fang,Liu,Ito,Xu}.
These new classes of 2D magnetic materials generate an interesting perspective for their possible applications in spintronics.
To elucidate the fundamental differences between 2D and 3D magnets in response to an external magnetic field, 
we start with a single domain
magnet in which the magnetization is spatially uniform across the 
sample. The single domain magnet is usually a building block for magnetic memory devices in which the
direction of the magnetization can be well controlled by either the magnetic field or the electric currents. The most elementary magnetic property of a single domain is its very simple magnetic hysteresis described by the classical Stoner-Wohlfarth (SW) model \cite{Stoner,Tannous}, whose magnetic energy is 
\begin{equation}
E_{sw}=-K ({\bf \hat{z} \cdot M} )^2- {\bf H}\cdot {\bf M}
\end{equation}
where ${\bf M}$ is the magnetization vector, $\hat{\bf z}$ is the anisotropy axis with anisotropy energy $K$, and
${\bf H}$ is the applied magnetic field. The above simple SW model immediately gives rise to the well-known hysteresis loops for the different directions of the applied magnetic field, as shown in Fig.~(1) (a-d). 

In this paper, we study the magnetic hysteresis of two-dimensional single-domain magnetic particles with uniaxial anisotropy, i.e., the 2D Stoner-Wohlfarth (SW) model. Why does the above successful SW model for the conventional 3D magnet fail for 2D magnets? In 3D, the magnitude of the magnetization $ M_s (T)= |{\bf M}|$
is controlled by the exchange interaction between the neighboring spins and thus it weakly depends on the magnetic field or the magnetic anisotropy for temperatures sufficiently less than the Curie temperature. Since the hysteresis is measured with a constant temperature,
$M_s$ does not change for the entire range of the field in the hysteresis. In 2D, however, the magnitude of the magnetization depends on both the exchange interaction and the total effective field ${\bf H}_{eff}$ (the sum of the anisotropy and the applied field).  If the total effective field is zero, the magnitude of the magnetization would be zero; this is known as the Wigner and Mermin theorem \cite{Mermin}. The dependence of the magnetization on the total effective field is due to fundamentally strong spin fluctuation in low dimensions in which the number of low-energy excitations (long-wavelength magnons) diverges, i.e., the long-range ordering disappears. In the hysteresis loop, when the magnetic field is reversed to the opposite direction
of the magnetization, the total effective field becomes small and thus the magnetization reduces.  To describe the variation of 
both magnitude and direction of the magnetization
with the applied field, we use the self-consistent spin-wave method, which is equivalent to the random phase approximation
\cite{Tang}, to model the magnitude of the magnetization. Let us first show the 2D hysteresis along with the above 3D hysteresis
in Fig.~(1) (e-h) followed by our detailed theory and calculation in the next Section.

\begin{figure}
    \centering
    \includegraphics[width=8.5 cm]{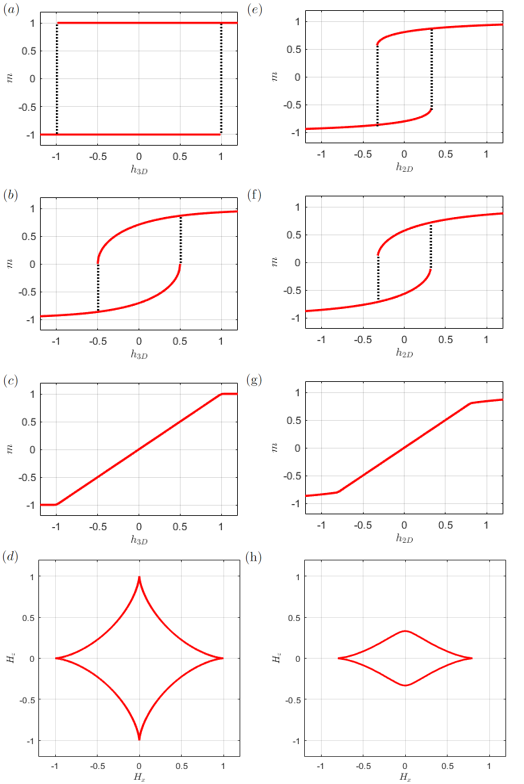}
    \caption{Magnetic hysteresis and switching Asteroid for the 3D (four left panels a-d) and 2D (four right panels e-h) Stoner-Wohlfarth model (where $m \equiv M/M_s$ , $h_{3D} \equiv H/2K$ and $h_{2D} \equiv H/2zAM_s$). $(a)$, $(b)$, and $(c)$, The hysteresis loops of a 3D magnet with the angles between the magnetic fields and the anisotropy axis at $0^{\circ}$, $45^{\circ}$, and $90^{\circ}$. $(d)$ The 3D Asteroid figure for magnetization switching fields. $(e)$, $(f)$, and $(g)$, are the hysteresis loops of the 2D magnet with the angles between the magnetic fields and the anisotropy axis at $0^{\circ}$, $45^{\circ}$, and $90^{\circ}$ at $T=0.6T_c$. $(h)$ The 2D Asteroid figure at $T=0.6T_c$.}
    \label{Figure 1}
\end{figure}
\section{Model}
The quantum version of the 2D SW model is 
\begin{equation}
    {\hat{\cal H}}=-J \sum_{<i,j>}{\hat{\bf S}}_{i} \cdot {\hat{\bf S}}_{j}- A \sum_{<i,j>} \hat{S}_{i}^z \hat{S}_{j}^z-\sum_{i}{\bf H} \cdot \hat{\bf S}_{i}
\end{equation}
where ${\hat{\bf S}}_{i}$ and $\hat{S}_{i}^z$ respectively are the spin and the $z$-component (taken as perpendicular to
the two-dimensional plane) of the spin operators at lattice site ${\bf R}_{i}$, $J$ is the isotropic exchange integral, $A$ is the anisotropic exchange integral (it is worth mentioning here that the anisotropy energy in the 3D classical limit $K$ is equivalent to $zAM_s$ in the quantum model, where z is the number of the nearest-neighbour sites), $<ij> $ indicates the sum over nearest neighbors, and $ {\bf H}$ is the external field. To 
determine the magnetization, we have developed a random phase approximation (RPA) in which the transverse spin fluctuation
and the longitudinal spin fluctuation are decoupled, and we have arrived at the self-consistent equation for the magnetization
\cite{Tang},
\begin{eqnarray}
M=M_s -\int \frac{d^2k}{(2\pi)^2} \frac{2M}{e^{\beta E_k}-1}
\end{eqnarray}
where $M_s$ is the magnetization at zero temperature, and $E_k$ is the magnon energy; in the long wave length limit,
$E_k =zM( J k^2/2 + 2A) + H$ (assuming the field is along the direction of the anisotropy field). Eq.~(3) has a straightforward 
explanation: the magnetization is subtracted by the number of the magnons which are softened by the factor of $M$ at the finite 
temperature. We note that a) Eq.~(3) is the RPA approximation for spin-1/2; the higher spins would lead to a more complicated
equation; b) the RPA is considered an excellent approximation for temperature sufficiently lower than the Curie 
temperature, and c) we consider the magnetic anisotropy from the anisotropic exchange rather than on-site anisotropy in the form of $-A(S_i^z )^2 $. By using the quadratic dispersion in the energy,  we may integrate out $d^2k$, resulting to a simple analytical expression,
\begin{equation}
M=M_s -\frac{1}{ \pi zJ} \left( \frac{1}{\beta} \ln \left| \frac{e^{\beta (\Delta +W)}-1}{e^{\beta \Delta}-1} \right| -W \right)
\end{equation}
where $\Delta = 2zAM +H$ and $W=2\pi zJM$ are the effective magnon gap and the magnon bandwidth, respectively. Clearly, the magnetization depends on the magnetic field even for temperature significantly lower than the Curie temperature. 
\section{External field in the direction of the anisotropy}
\begin{figure}
    \centering
    \includegraphics[width=8.5 cm]{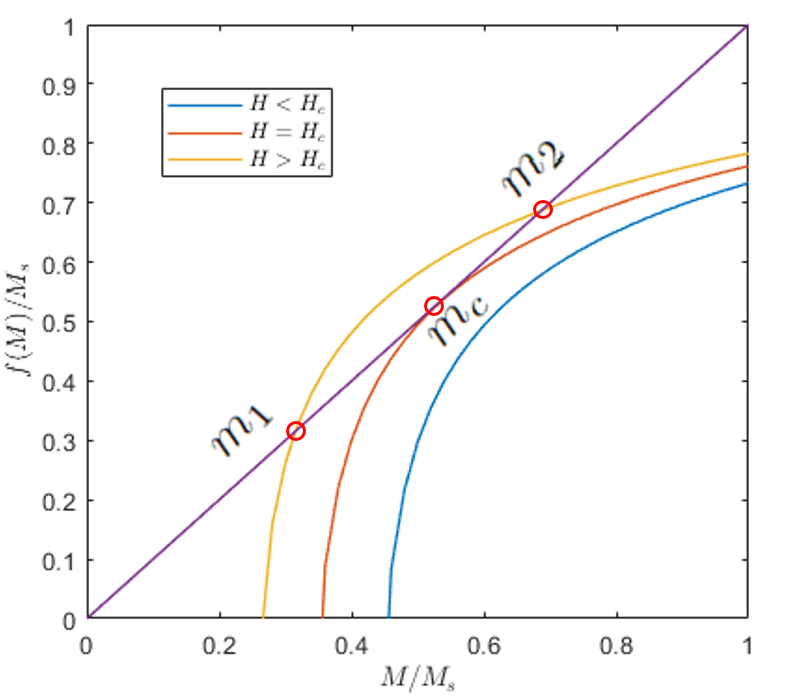}
    \caption{The solutions of Eq.~(4) for three different magnetic fields. For the field less than $|H|< H_c$ the equation has two solutions $m_1$ and $m_2$. At $H= -H_c$ (the coercive field) the equation has one solution which is the critical magnetization at which the magnetization reversal occurs. }\label{fig:fig.1}
    
\end{figure}
\begin{figure}
    \centering
    \includegraphics[width= 8.5 cm]{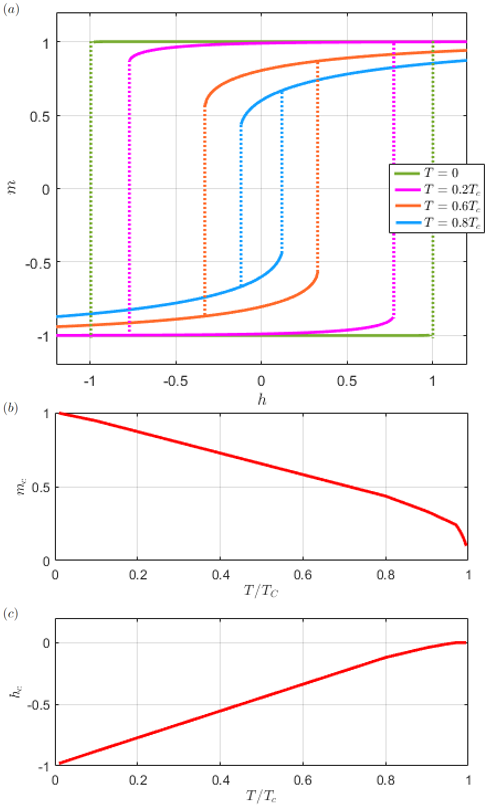}
    \caption{$(a)$ The hysteresis of a single domain 2D magnet at different temperatures ($2zAM_s=1$). $(b)$ The temperature dependence of the critical magnetization. $(c)$ The temperature dependence of the coercive field.}
    \label{fig:fig.1}
\end{figure}
The numerical solution of $M$ for the field in the direction of the anisotropy field is readily solved from Eq.~(4). A simple 
way to obtain a solution for given parameters (temperature, field, and anisotropy) is to plot two functions $y=M $ and $y=f(M)$ where $f(M)$ is the right side of Eq.~(4). Note that the function $f(M)$ is only physically meaningful when the magnon energy
gap is positive, i.e., $ \Delta > 0$. The negative gap is unstable such that the magnetization reversal takes place.
In Fig.~(2), we show $f(M)$ for three different magnetic fields: for a positive or small negative field,
$M=f(M)$ has two solutions, representing an energy minimum (the solution with a larger $M$) and an energy maximum. At a critical negative 
magnetic field, there is only one solution, which is also known as the coercive field. Beyond the critical field, there is no 
solution for $M>0$, indicating magnetization reversal occurs. 

Compared with the conventional SW model, the hysteresis shown in Fig.~(1e) is no longer square.
The reduction of the magnetization near the critical value of the field is caused by the reduced effective gap and thus the
increased number of magnons. Since the magnon population depends on the temperature, the
magnetization at the critical magnetic field decreases significantly at higher temperatures as shown in Fig.~(3). This contrasts with the 3D magnet, which is essentially independent of temperature.
\begin{figure}
    \centering
    \includegraphics[width=8.5 cm]{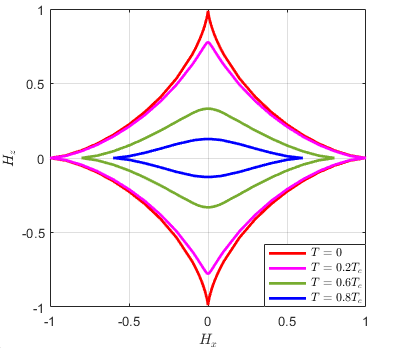}
    \caption{The temperature dependence of the Asteroid diagram of 2D single domain magnet ($2zAM_s=1$).}
    \label{fig:fig.1}
\end{figure}
\section{External field at an arbitrary direction}
We now consider the hysteresis loop with the field in an arbitrary direction, ${\bf H} =H_0 ( \hat{\bf z} \cos\theta + \hat{\bf x} \sin\theta )$ where $\theta$ is the angle between the applied field and the ${\bf z}$-axis . The total effective field is the sum
of the external and anisotropic field, ${\bf H}_{t} = {\bf H} + 2zA ({\bf M} \cdot {\hat{\bf z}}) \hat{\bf z}$ where the direction and the magntitude of ${\bf M}$ need to be self-consistently determined.  At equilibrium, the magnetization ${\bf M}$ is always parallel to $H_{t}$, i.e.,
\begin{equation}
\frac{M_x}{M_z }=\frac{H_0 \sin\theta }{H_0 \cos\theta + 2zA M_z}
\end{equation}
Eq.~(4) remains valid as long as the magnon gap is replaced by the total magnetic field in the direction of the magnetization,
i.e., $\Delta = {\bf H}_{t} \cdot {\bf M}/M $. Thus, Equations (4) and (5) determine the magnetization for arbitrary direction
of the magnetic field. As an example, we show in Fig.~(1f) the hysteresis for the field direction at $\theta = 45^{\circ}$. Comparing
3D, Fig.~(1b), and 2D, Fig.~(1f), the coercivity is smaller while the magnetization jumps at a positive value. For $\theta =90^{\circ}$, i.e., the hard axis loops, both 2D and 3D hysteresis are single valued. However, the 2D SW model has a non-zero slope
even above the anisotropy field while the 3D SW would be completely saturated above the anisotropy field.

Finally, we construct the critical values of the magnetic field for the magnetization reversal for all directions of the magnetic field, known as the Asteroid figure. When the magnetic field increases across the Asteroid line,
the reversal occurs. In the classical SW model, the Asteroid line can be readily derived from Eq.~(1) and the analytic 
expression of the Asteroid is $H_z^{2/3} + H_x^{2/3} = (2K)^{2/3}$. In 2D SW model, the Asteroid is highly temperature
dependent as shown in Fig.~(4). At low temperature, the Asteroid figure resembles that of the 3D SW model. At higher
temperature, the magnetization reversal for the longitudinal field (parallel to the anisotropy field) is more effective than
for the transverse field; this is because for the same magnitude of the field, the longitudinal direction reduces the magnon
gap more than the transverse direction, leading to the asymmetry of the Asteroid figure in the direction of the applied field.

In summary, we have formulated the magnetic hysteresis loops of 2D single domain magnets. Compared to 3D Stoner-Wohlfarth
single domain model, the 2D magnetic hysteresis is more complicated due to the fundamentally stronger spin fluctuations. 
We have used the equilibrium magnetization formulation which has been derived previously by the random phase approximation.
As long as the temperature is not too close to the Curie temperature, the RPA provides an excellent approximation.

This work was partially supported by the U.S. National Science Foundation under Grant No. ECCS-2011331.


\end{document}